# A hierarchical approach to deep learning and its application to tomographic reconstruction

Lin Fu and Bruno De Man

GE Research, Niskayuna, NY 12309, USA

*Abstract*

Deep learning (DL) has shown unprecedented performance for many image analysis and image enhancement tasks. Yet, solving large-scale inverse problems like tomographic reconstruction remains challenging for DL. These problems involve non-local and space-variant integral transforms between the input and output domains, for which no efficient neural network models have been found. A prior attempt to solve such problems with supervised learning relied on a brute-force fully connected network and applied it to reconstruction for a $128^4$ system matrix size. This cannot practically scale to realistic data sizes such as $512^4$ and $512^6$ for three-dimensional data sets. Here we present a novel framework to solve such problems with deep learning by casting the original problem as a continuum of intermediate representations between the input and output data. The original problem is broken down into a sequence of simpler transformations that can be well mapped onto an efficient hierarchical network architecture, with exponentially fewer parameters than a generic network would need. We applied the approach to computed tomography (CT) image reconstruction for a $512^4$ system matrix size. To our knowledge, this enabled the first data-driven DL solver for full-size CT reconstruction without relying on the structure of direct (analytical) or iterative (numerical) inversion techniques. The proposed approach is applicable to other imaging problems such as emission and magnetic resonance reconstruction. More broadly, hierarchical DL opens the door to a new class of solvers for general inverse problems, which could potentially lead to improved signal-to-noise ratio, spatial resolution and computational efficiency in various areas.

**Keywords**: computed tomography, image reconstruction, deep learning, hierarchical.

## 1. Introduction

The surge in deep learning (DL) imaging research in recent years has resulted in a plethora of applications and network architectures [1–4]. Most of these approaches can be categorized in two major areas:

*Image analysis* applications seek to make a decision or diagnosis. The input to the DL network is an image and the output is a discrete set of labels (Fig. 1a). All the voxels in the input image are indirectly linked to the final labels through a complex neural relationship. Examples of this category include the classification of images as cats and dogs [5] and the diagnosis of medical images as malignant or benign [6].

*Image enhancement* applications aim to improve the image in some context-dependent way. The input is an image; the output is another image (Fig. 1b). Input and output images are spatially 'linked' so the network exhibits a high degree of locality – input values mostly affect only output values in their immediate vicinity – hence training is commonly performed with small image patches. In this category, typical applications include image sharpening [7], image denoising [8–10], and semantic segmentation [11] [12].

The focus of this paper will be a third category of applications, *domain transforms*. In these problems both input and output are images (or relatively large datasets represented as two- or higher-dimensional arrays), but without any direct spatial linkage between the two (Fig. 1c). This is the case for many inverse problems. The transform between the two domains may take the form of an integral transform and the spatial relationship between the input and output is often non-local and shift-variant - each input image voxel contributes to all output image voxels and – vice versa – each output image voxel is defined by all input image voxels. This category includes the Fourier transform [13], tomographic reconstruction[14], and many inverse problems in geoscience, physics, healthcare, and defense applications.

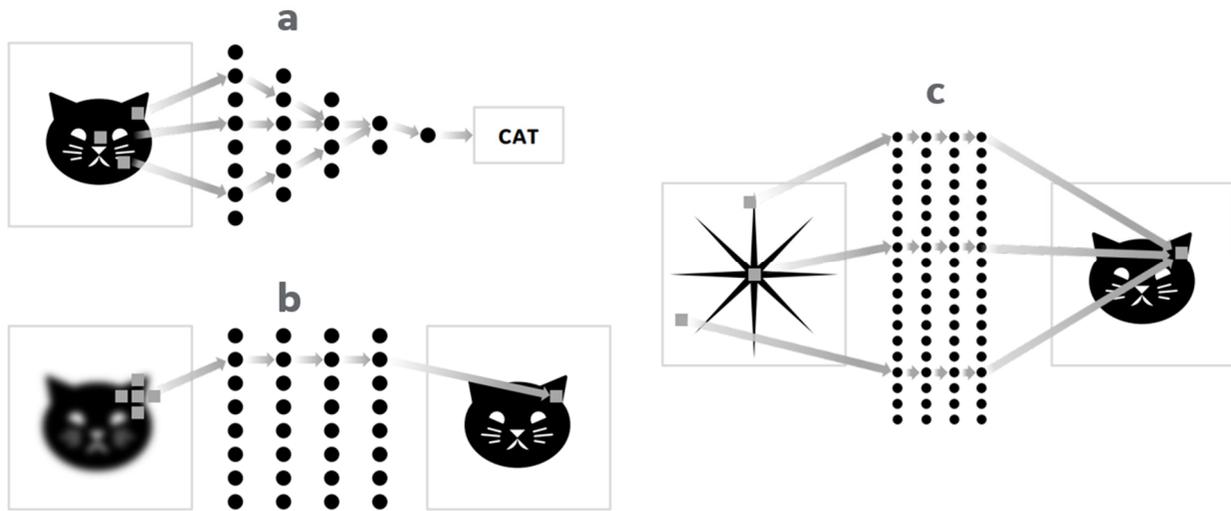

*Figure 1: Schematic representation of deep learning for three categories of problems: image analysis (a), image enhancement (b) and domain transforms (such as Fourier transform) (c). Domain transform remains challenging for deep learning because there is no direct spatial linkage between the input and output domains, hence large-scale problems are usually infeasible to a neural network.*

The challenge with this third category is the very high dimensionality of the neural networks. Unlike the other two categories of problems, it can be difficult and infeasible to train a network for the third category applications. In matrix terminology, the absence of simple spatial linkage results in non-sparse matrices or non-convolutional matrices, making conventional neural network models less effective. A brute-force network model of such problems would result in an extremely high-dimensional network, which poorly generalizes to larger scale problems. For example, using a dense network to model the transformation between two-dimensional images of size 512x512 would result in a number of network weights on the order of $512^4 \approx 64$ billion. This also explains why prior attempts to address the third category of problems using DL are typically limited to 64x64 or 128x128 images [13] [14].

Here we present a scalable DL approach for the third category of problems along with a more broadly applicable hierarchical DL framework. The proposed approach decomposes a high-dimensional transform into hierarchical stages, where the individual stages employ only local transforms that can be efficiently mapped to neural networks. Such hierarchical decomposition also leads to exponentially fewer parameters than a generic network would need. In matrix terminology, this is analogous to approximation of a dense matrix by a product of sparse matrices. In addition, we introduce the idea of training DL networks with computer simulated random noise patterns, which overcome the needs for large amount of training data for supervised learning.

More specifically, we apply the hierarchical reconstruction to full-size two-dimensional (2D) computed tomography (CT). Extension of the general hierarchical DL framework to three dimensions and to other applications such as magnetic resonance reconstruction and emission reconstruction is conceptually straightforward but requires future research. To our knowledge all previous CT reconstruction algorithms ever developed can be categorized as direct reconstruction or iterative reconstruction. Direct reconstruction approaches such as filtered backprojection (FBP) [15] [16] [17] have relatively low complexity and result in perfect images under strongly idealized conditions. Iterative reconstruction approaches [18] [19] [20] can more effectively deal with noise and other non-idealities but have high computational complexity. The application of DL to image reconstruction opens the door to an entirely new third type of reconstruction. However, due to the challenges associated with domain transforms, the only prior work attempting to solve this problem is AUTOMAP[14], which relied on dense networks that cannot scale to full-size data. The approach we present here was conceived [21] independently of the AUTOMAP approach and has the advantage of being fundamentally scalable to full-size problems. In several other contexts of CT imaging, DL-based signal processing components have been introduced to augment conventional direct or iterative CT reconstruction algorithms [22], but these typically rely on the structure of a conventional algorithm, including accelerating iterative reconstruction [23], post-reconstruction image denoising or restoration [8] [9] [10] [24], unrolled iterative reconstruction[4] [25], optimization of projection or image-domain filter weights [26] [27], analyzing raw data and by-passing reconstruction entirely [28]. These efforts still have not addressed the problem of the full domain transform from the perspective of data-driven supervised learning.

It is worth noting that the proposed hierarchical reconstruction framework is not comparable to the previously published hierarchical projection and backprojection approach[29]. The latter is a fast, numerical implementation of a traditional component and is used as part of a direct or iterative reconstruction algorithm. The hierarchical DL framework proposed here is a radical change that does not rely on an analytic inversion or on an iterative data fit optimization. In other words, direct reconstruction relies on a mathematical inversion of a Radon transform or a cone-beam transform, and iterative reconstruction relies on numerical inversion of the same. We propose and demonstrate an approach that relies on a data-driven learnt inversion.

## 2. Methods

### 2.1 General theory

Consider inverse problems whose forward models can be written in the form of the Fredholm integral equation of the first kind

$$p(v) = \int K(u,v) f(u) du. \tag{1}$$

The problem is given the kernel function $K(u,v)$ and the observation $p(v)$, to infer the function $f(u)$. When the kernel $K(u,v)$ is non-local and space-variant, it is difficult or impossible to model the solution with a neural network, especially for large-scale problems.

A complicated transform can sometimes be decomposed into a sequence of simpler hierarchical steps, providing an avenue for fast and efficient algorithms. Here we propose a new framework to make such problems more amenable to data-driven supervised learning. Since the root cause of the problem are the non-local kernels, we introduce a virtual intermediate data domain by applying a window function $\Pi(\cdot)$ to the integration

$$q_\alpha(v,t) = \int K(u,v) f(u) \Pi\left(\frac{t-u}{\alpha}\right) du,$$

where $\Pi(x) = 1$ for $|x| < 0.5$ and $\Pi(x) = 0$ elsewhere, $\alpha$ is a scale factor controlling the window size, and $t$ is a parameter specifying the window location. By varying $\alpha$, a continuum of intermediate representations between $p(v)$ and $f(u)$ (like a homotopy) can be obtained. Both $p(v)$ and $f(u)$ can be viewed as marginal cases of $q_\alpha(v,t)$. In the limiting case that $\alpha \to 0$, the window becomes a Dirac delta impulse, thus $\lim_{\alpha \to 0} q_\alpha(v,t) = K(t,v) f(t)$ and the integral vanishes. On the other hand, a larger $\alpha$ corresponds a wider integration window, making $q_\alpha(v,t)$ closer the original measurement $p(v)$.

Although $q_\alpha(v,t)$ represents virtual data that cannot be physically measured, we propose that they could be synthesized and serve as a form of intermediate training labels. With progressively smaller integration window, $\alpha_1 > \alpha_2 > \cdots > \alpha_N$, the original problem may be decomposed into a series of incremental steps

$$p(v) \to h_{\alpha_1}(v,t) \to h_{\alpha_2}(v,t) \to \cdots \to h_{\alpha_N}(v,t) \to f(u).$$

$q_\alpha(v,t)$ represents a continuum between $p(v)$ and $f(u)$, and the incremental transitions from $h_{\alpha_n}(v,t)$ to $h_{\alpha_{n+1}}(v,t)$ possess a high degree of locality like a conventional image enhancement task. This gives rise to a hierarchical DL architecture that can tackle original domain transform problems by breaking them down to a hierarchy of simpler problems that are well-suited for implementation with neural networks.

### 2.2 Hierarchical CT reconstruction

Here we present a detailed example of applying the hierarchical framework for solving domain transforms to the problem of CT reconstruction. The purpose of CT reconstruction is to infer an image of the internal structure of an object from projection measurements taken along various rays passing through the object. The CT forward model can be expressed in the form of the Fredholm equation, with the kernel $K(u,v) = \delta(u_x \cos v_\theta + u_y \sin v_\theta - v_r)$, where $\delta(\cdot)$ is the Dirac delta function. The kernel represents projection rays parameterized by an offset $r$ and an angle $\theta$ in a 2D plane.

We define the three data domains for hierarchical CT reconstruction (Fig 2). First, the original CT measurements $p(r, \theta)$, called the projections or sinogram, reside in the **line integral domain**. This represents line integrals with the original kernel over the entire length of the object. We introduce a coordinate system $x - y$ and a second coordinate system $r - t$, rotated by $\theta$. Then the projection lines are parametrized by a radial distance $r$ and a rotation angle $\theta$, and in the three-dimensional (3D) case, also by a $z$-distance (or cone angle). There is no depth resolution along $t$ since the line integrals are over the entire projection lines. Second, the reconstructed image $f(x, y)$ is represented in the regular **voxel domain.** A voxel value can be interpreted as a marginal case of a line integral, where the integral is only over the length of the voxel. The voxel values are parametrized by $x$- and $y$-coordinates or by $r$- and $t$-coordinates, but there is no angular parameter because all voxels share the same angle; and in the 3D case also by a $z$-coordinate. Finally, the proposed intermediate data $q_\alpha(r, \theta, t)$ are defined in the **partial line integral domain**, generated by restricting the line integral with a window function $\Pi(\frac{-u_x \sin\theta + v_y \cos\theta - t}{\alpha})$ which corresponds to a line segment of length $\alpha$ centered at position $t$ along the projection line. The partial line integrals are parameterized by a radial distance $r$, an in-plane rotation angle $\theta$ (but typically fewer rotation angles than in the line integral domain), and a depth $t$. The partial line integral is also associated with the parameter $\alpha$, which specifies the width of the window function. Both the line integrals and the reconstructed voxels can be viewed as marginal cases of the partial line integrals.

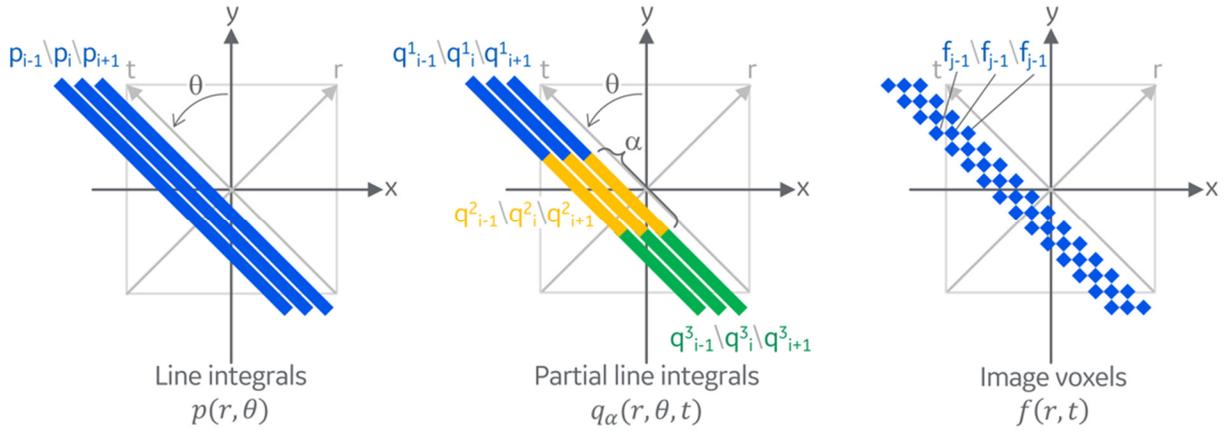

*Fig. 2. Partial line integrals are proposed as an intermediate representation between line integrals and image voxels, both of which being its marginal cases. The original CT measurements are line integrals over the entire length of the object. The reconstructed image is represented by the image voxels, which are essentially line integrals over the length of the voxel (except for a typical scale factor normalizing by the voxel size).*

With the intermediate domain defined, CT reconstruction may be viewed as a process that progressively solves for partial line integrals with shorter and shorter integration length, until regular image voxels are reached.

$$p(r,\theta) \to q_{\alpha_1}(r,\theta,t) \to q_{\alpha_2}(r,\theta,t) \to \cdots \to q_{\alpha_N}(r,\theta,t) \to f(r,t)|_{\theta=0}$$

where $\alpha_1 > \alpha_2 > \cdots > \alpha_N$. This decomposition into incremental stages gives rise to a novel hierarchical flow for CT reconstruction (Fig 3). The input data are line integrals with various offsets and orientations, but without depth resolution. Then, the proposed network transforms line integrals into partial line integrals, gaining depth resolution, while the number of angular sampling is reduced. Multiple intermediate stages with progressively finer depth resolution may be used, although only a single intermediate stage is shown in Fig 3. Finally, the partial line integrals are transformed into regular image voxels as output, further gaining depth resolution. Overall, as data go through the hierarchy, the depth resolution increases, while the angular sampling density decreases, keeping the total amount of data unchanged. Such a reconstruction may be viewed as a process that gradually trades angular resolution in the original sinogram for better depth resolution in a partial line integral domain and ultimately, generates the reconstructed image. Intuitively, the incremental elementary reconstruction step in hierarchical reconstruction is analogous to a limited-angle tomosynthesis reconstruction, where a coarse level of depth resolution can be estimated from only a few projections in adjacent angles. In hierarchical reconstruction such incremental reconstruction step is repeated to incorporate information from wider and wider angular ranges and ultimately produces a regular reconstructed image with isotropic spatial resolution.

The key benefit of this hierarchical framework is that the elementary reconstruction from one stage to the next is localized in terms of number of angular ranges, number of radial distances and number of depth positions considered. The estimation of the partial line integrals only requires the line integrals that are at nearby angular positions and at nearby radial positions. Similarly, the estimation of the voxel values requires as inputs only the partial line integrals that are at nearby radial and depth positions (indicated by red rectangles in Fig. 3). Hence the hierarchical reconstruction effectively factorizes a non-local operator as a product of a series of local (hence sparse) operators, making the algorithm suitable for efficient implementation with a neural network.

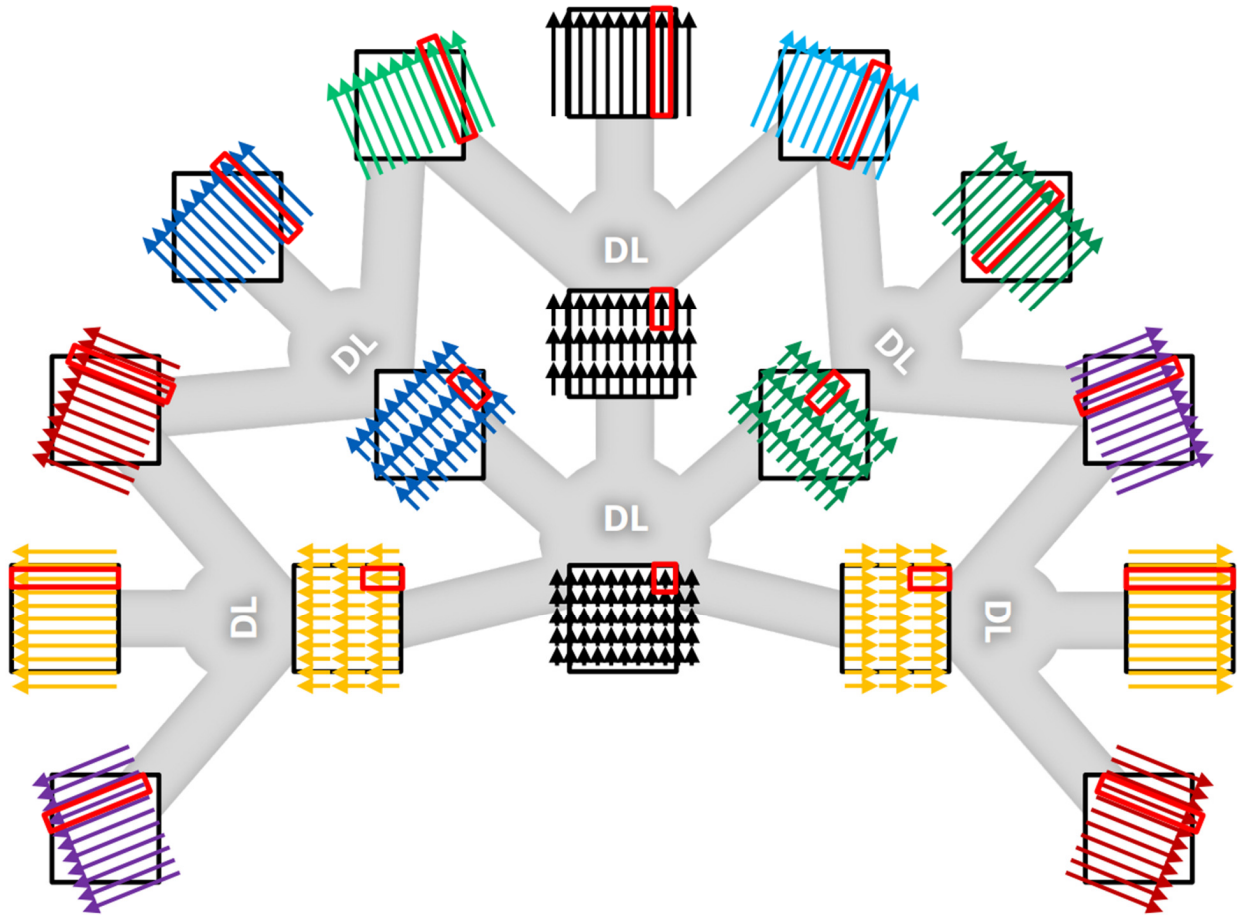

*Fig. 3. Illustration of the concept of hierarchical CT reconstruction. The diagrams in the outer ring represent the input data, i.e., line integrals at various rotation angles and radial offsets. The diagrams in the middle ring represent data in the intermediate domain, i.e., partial line integrals. The diagram at the center represents the output, i.e., the reconstructed image. The gray connections illustrate the flow of data through the network. As data go through the hierarchy, the depth resolution improves, while the number of angles decreases. The final reconstructed image is formed when the number of depth bins equals the desired size of the reconstructed image, and the number of angles reaches unity. The key benefit of hierarchical reconstruction is that the operations in each hierarchical stage are relatively localized, making it suitable for efficient implementation as neural networks. The red rectangles illustrate the localized correspondence across hierarchical stages.*

Fig 4 shows the actual outputs from a hierarchical reconstruction network at different hierarchical stages for a realistic CT reconstruction example. (More details about the network architecture and training are provided in the next sections). The intermediate reconstructions are outputs from hidden layers and illustrate the inner workings of the hierarchical network. The original input sinogram gradually transforms into the final reconstructed image through a number of intermediate representations. The intermediate images have non-isotropic voxel sizes with coarser resolution along the depth dimension ($t$) compared to the radial channel dimension ($r$). As reconstruction progresses, the intermediate images become more isotropic and the number projection angles (or the number of partial images) decreases, until the final reconstructed image is formed.

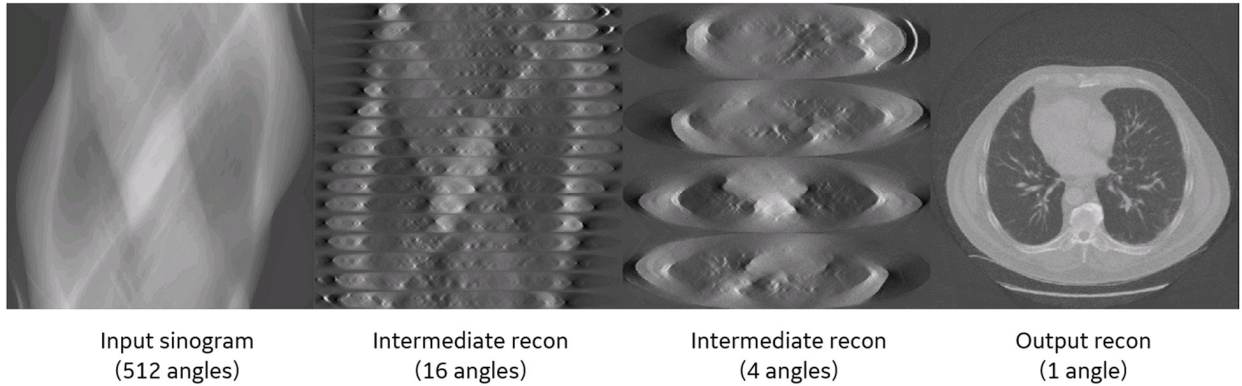

Fig. 4. The progression of reconstruction through hierarchical stages. Two intermediate reconstructions are shown. All partial images at different rotation angles are stacked vertically for better visualization.

### 2.3 Network structure

The results in Fig 4 were generated by a feed forward network consisting of 5 hierarchical stages ($L1$ to $L5$). The overall network structure is illustrated in Fig. 5. The input sinogram contains 512 parallel-beam CT detector channels and 512 projection angles equally spaced over 360 degrees. The output image size is 512x512. As the reconstruction progresses through the hierarchy ($L1$ through $L5$), the number of angular bins is reduced, and the number of depth bins is increased. The data dimension at each hierarchical stage is shown in Table 1.

As mentioned earlier, the operation in each hierarchical stage is relatively localized. To implement this efficiently, we use "sparse connection" network layers, where non-zero network weights are stored in a sparse matrix. In the limiting case where the matrix had no zero elements, the sparse connection layer would become a fully connected layer. We did not enforce rotational symmetry or other constraints in the non-zero network weights, making the sparse layers very flexible in expressing more general operations. It is worth noting that the sparse connection layer is different from a drop out layer, which removes output neurons instead of the connections between the input and output neurons of this layer.

Let $n$ denote the data dimension of the sparse connection layer, in this case $n = 512 \times 512$. Because an input neuron only affects a relatively small neighborhood of output neurons, the number of network parameters in each sparse connection layer is on the order of $O(n)$. Because the number of angles decreases exponentially from one hierarchical stage to the next, the total number of hierarchical stages is expected to be on the order of $O(\log n)$. Multiplying these two factors, a hierarchical reconstruction network overall would require a number of network parameters on the order of $O(n \log n)$. This contrasts with a generic fully connected network, which would require $O(n^2)$ parameters. That would become intractable for realistic data sizes. In our implementation, all five sparse connection layers in total used about 42 million trainable parameters, only 0.06% of those in a fully connected layer of the same input and output dimensions ($512^4$ or 69 billion parameters). For the 3D case, the number of parameters of a fully connected network would be on the order of $512^6$ or 18 quadrillion, where we estimate – by extrapolation – the proposed approach would require on the order of 21 billion trainable parameters.

In this initial study, the network layer at stages $L1$ through $L5$ used linear activation. The first layer was a sinogram domain convolutional layer with a filter kernel size of 512. The last four layers were image domain convolution layers with 3x3 kernels with ReLU activation. Optional convolutional layers and nonlinear activation could also be inserted between hierarchical stages, which could be a topic for a follow up study.

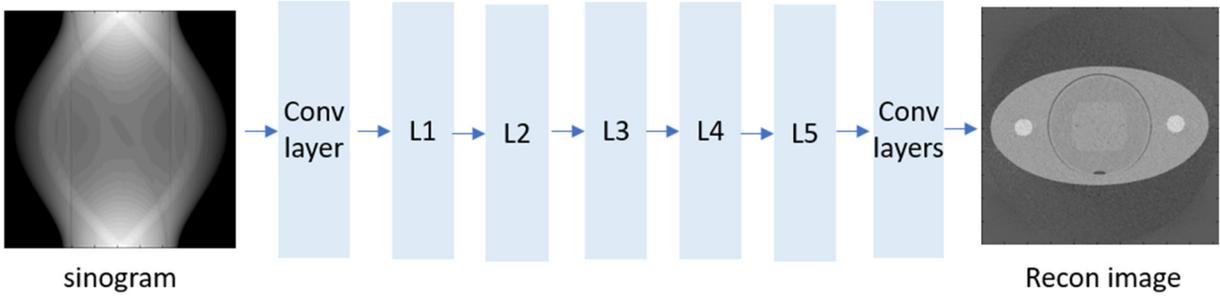

*Fig. 5. Network architecture for hierarchical CT reconstruction. Layers L1 through L5 are sparse connection layers that gradually transform the data from the sinogram domain to the image domain. Optional sinogram-, image, or intermediate-domain CNNs can be incorporated to further optimize image quality.*

*Tab. 1. Data dimension at different hierarchical stages*

|  | Data dimension | |
| --- | --- | --- |
| Hierarchical Stage (L) | #depth bins ($N_t$) | #projection angles ($N_\theta$) |
| L=0 (line integrals) | 1 | 512 |
| L=1 | 2 | 256 |
| L=2 | 8 | 64 |
| L=3 | 32 | 16 |
| L=4 | 128 | 4 |
| L=5 (isotropic resolution) | 512 | 1 |

## 2.4 Training

Initial training of the hierarchical network was performed with 200 realization of computer generated white Gaussian random noise patterns and their corresponding distance-driven[30] forward projections. These data pairs encode the tomographic transform and thus can be used for training the inverse transform. We analytically generated the training labels for the intermediate representation levels (partial line integrals) by FBP reconstruction on non-isotropic voxel grids corresponding to the data dimension at each hierarchical stage *L*. (Alternatively, these intermediate datasets could be computed by reprojecting the noise images over partial line integrals). An instance of the training labels is shown in Fig. 6.

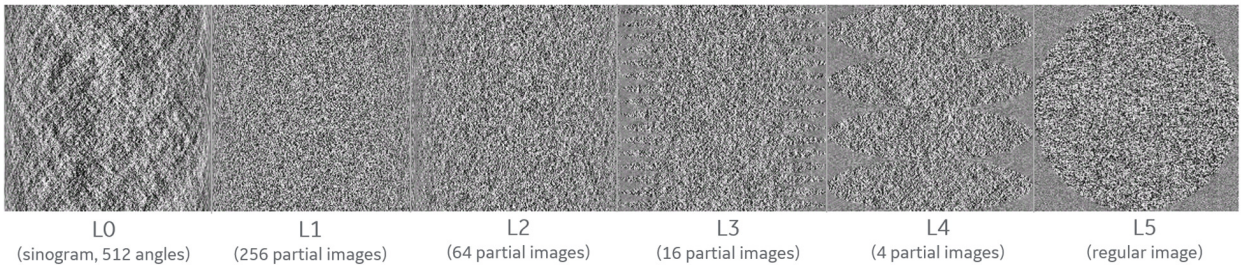

L0 (sinogram, 512 angles)   L1 (256 partial images)   L2 (64 partial images)   L3 (16 partial images)   L4 (4 partial images)   L5 (regular image)

*Fig 6. An instance of the noise pattern realizations used for training labels for intermediate representation levels. For better visualization, partial reconstructions of the same hierarchical stage (for different angles) are stacked together vertically.*

Each sparse connection layer was first pre-trained individually using the training labels at the corresponding hierarchical stage. Only the non-zero weights within the sparse connection layer were updated during the training. After the pre-training, an end-to-end training of the entire network was performed (i.e.: without intermediate training labels). When trained end-to-end, to address the issue of vanishing gradients, at each iteration we randomly chose a single network layer for update, while freezing the network parameters of the other layers. All non-linear activation units were disabled when trained with pure noise. The mean squared error was used as the training loss. Stochastic gradient descent with a batch size of 20 was used. The sparse connections were initialized with all ones and convolution kernels were initialized with the Glorot uniform initializer. All networks were implemented in TensorFlow with Keras frontend. Training was performed with a nVidia Tesla V100 GPU.

Finally, the network was further refined by an end-to-end training/validation with 200 clinical datasets, now with non-linear activation units enabled. The motivation for this selective non-linear activation strategy is based on the intuition that noise-based training is well-suited for learning the inverse transform, whereas clinical training datasets are better suited for also providing prior information on what high-quality, low-noise images look like and hence motivating (non-linear) denoising elements. Clinical training pairs were obtained by reprojecting clinical CT images and inserting random noise to emulate additional measurement noise. For testing, 50 additional clinical CT datasets were used, obtained the same way as the above training pairs, i.e.: by reprojection and noise insertion.

## 3. Results

Fig 7 shows the loss function during end-to-end training with pure noise. Good convergence behavior is observed for both the training and the validation data sets.

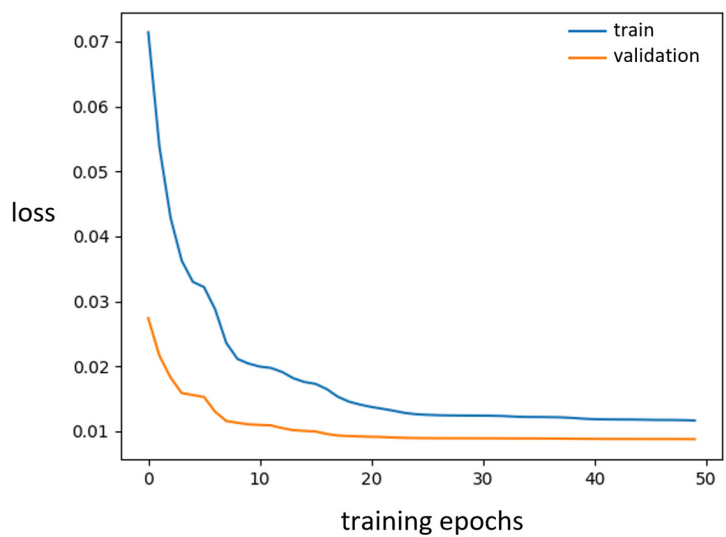

*Fig. 7. Loss function of the hierarchical network when trained end-to-end with pure noise.*

To visualize the network connections, an array of points was fed as the input to the sparse connection layer L5 (Fig 8 left). In the output of L5, each input point is mapped to a small line segments in various orientations (Fig 7 right). This illustrates the partial line integral representation at L4. This also illustrates that the network inputs and outputs have local-to-local correspondence.

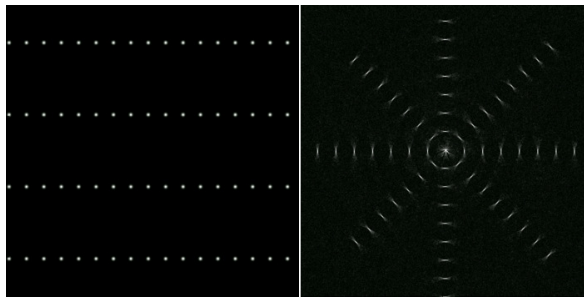

*Fig. 8. Visualization of the sparse network connection at L5. Four rows of points (left) are fed as input the sparse layer and are mapped to small line segments in four angular orientations at the output (right).*

Fig. 9 shows the final hierarchical DL reconstructions along with the corresponding FBP reconstructions and the original clinical images (prior to reprojection and noise insertion). The hierarchical network trained with pure noise produced reasonable reconstructions compared to FBP. The results obtained from a network with additional training with clinical data pairs further reduced reconstruction noise.

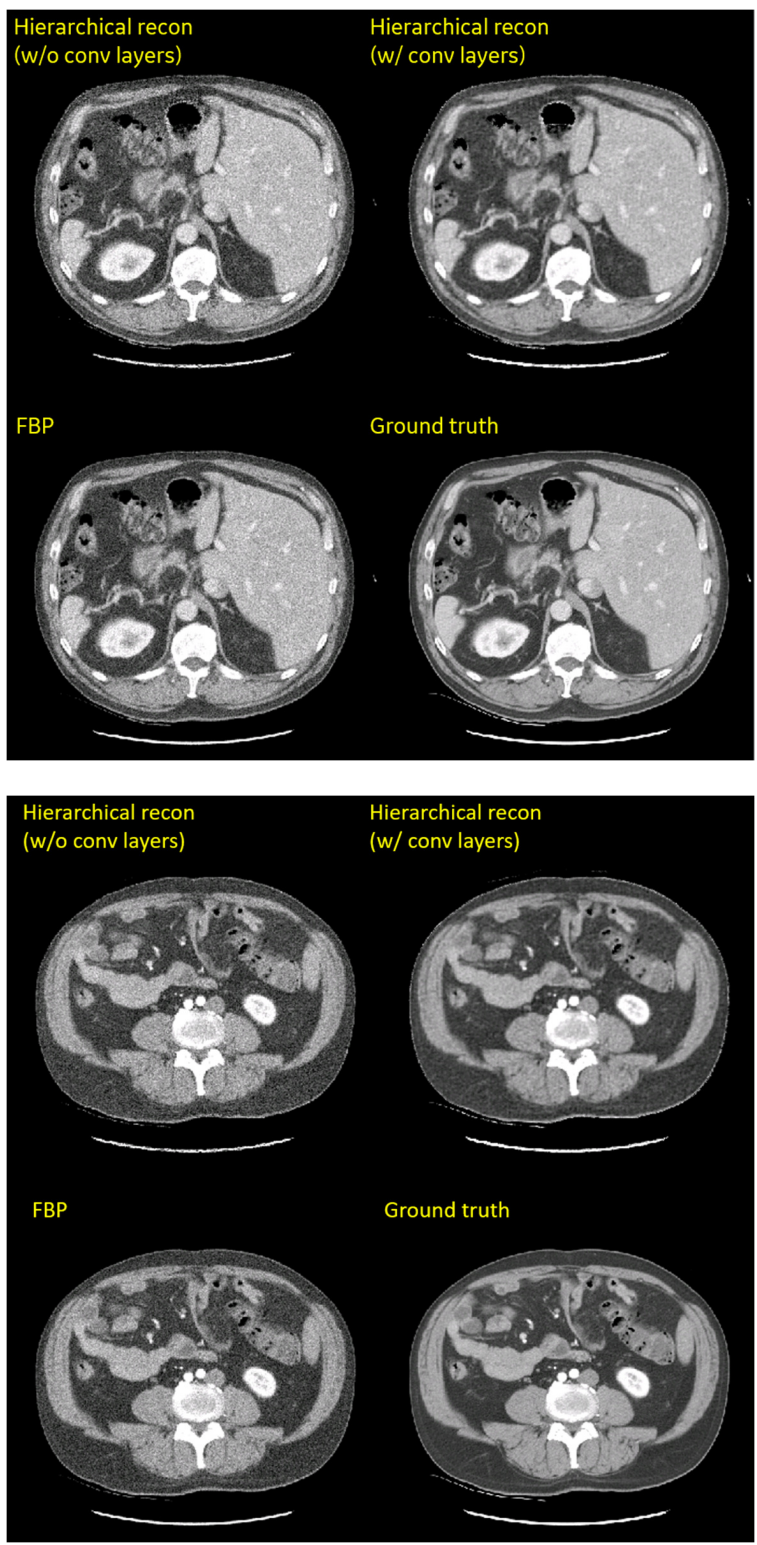

*Fig. 9. Examples images of hierarchical reconstruction in comparison with reference reconstructions. Two anatomical slices are shown. Window width/level = 400/0 HU.*

## 4. Discussion

The results illustrate the feasibility of hierarchical deep learning for CT reconstruction and also provide some intuition in its inner working. The hierarchical decomposition framework shows a progressive transformation from many view angles to fewer view angles with more and more depth resolution (Fig. 4). This illustrates the intuitive origin of the proposed approach, as humans also perceive some depth information from the limited stereoscopic range of the eyes [31]. Similarly, tomosynthesis imaging produces images with limited depth resolution from a limited angular view range [32]. The final stages in our proposed framework are more commensurate with time-of-flight (TOF) PET reconstruction, where measurements with some level of depth resolution (depending on the TOF timing resolution) provide higher quality reconstruction than traditional PET measurements[33].

This hierarchical framework effectively localizes each phase of the reconstruction in the sense that the reconstructed variables in each hierarchical level only depend on a limited range of variables in the adjacent hierarchical levels, as illustrated in Figure 8. Mathematically, this means that the respective system matrices are multiple orders of magnitude smaller than the system matrix of the full forward model, which we believe is the key to making learnt reconstruction possible for realistic data sizes. While our work so far was limited to (full-size) two-dimensional datasets, the extension to three-dimensions is conceptually straightforward. Since sparse connection layers are define purely based on correlation between hierarchical levels, extension to three dimensions only means an increase in data size, which our approach can afford given its sparsity level.

The image quality of these first full-scale learnt CT reconstruction results visually matches that of FBP reconstruction, a technique that has been the gold-standard for decades. The new approach has a long runway ahead in terms of possible refinements. We have no doubt that image quality can be vastly improved by building on the proposed hierarchical framework and incorporating more advanced network layers, connections and training schemes. In the future, it could potentially outperform state-of-the-art direct and iterative reconstruction techniques and combine the best of both other classes of reconstruction, i.e.: exceed the image quality of iterative reconstruction at computation times below those of direct reconstruction.

We introduced the idea of training DL networks with computer simulated random noise patterns, overcoming the needs for large amount of clinical training data for supervised learning. The rationale for using random noise images is rooted in the robustness of training the network without any pre-conceived notion of emphasizing certain locations, frequencies or patterns. The counter-argument is that this noise-based training does not teach the networks any prior information about what clinical images look like. In fact, it may steer the network towards reconstructing noisy patterns, which is not a desired outcome. In other words, the noise-training approach is well-suited for training a network to robustly perform the inverse Radon transform. The incorporation of prior information has proven to be highly powerful, initially through Markov Random Field regularization techniques [34], later in approaches such as dictionary learning [34] and non-local means [35], and most recently in deep learning based priors [1,36,37]. Hence, we expect performance can be greatly enhanced by relying entirely on a wide variety of high-quality clinical images and corresponding noisy sinograms. Future research will include more extensive training, validation and comparison with state-of-the-art iterative reconstruction and more traditional deep learning techniques.

A key ingredient for the implementation of the proposed hierarchical network is the sparse connection layers. These layers are a general tool for realizing transforms with a high degree of locality without using a fully-connected network. In this study the sparse connections were pre-determined by the imaging geometry. As a future research topic, dynamic routing algorithms could be potentially used to prune or create connections during training and further optimize these connections.

In this study, linear activation was used for the sparse connection layers. Performance could be further improved by using non-linear activation throughout the sparse connection layers. Moreover, convolutional layers could be inserted between the sparse connection layers to further improve the capability of the network.

While the hierarchical framework suggests straightforward rotational symmetries (Fig. 3), these were currently not yet explicitly exploited. In principal, deep learning network could be at one angle and redeployed at all angles. This should greatly improve training efficiency and further decrease network dimensionality. This may be achieved manually or through more implicit structural changes and will be an interesting area for future research.

The proposed hierarchical framework is not limited to applications in tomographic reconstruction. Fourier-related transforms are another example that can be efficiently mapped to a neural network with the proposed hierarchical decomposition. To show this, one can express Fourier transform in the form of the Fredholm equation, with the integration kernel being $K(u,v) = e^{-iuv}$. As shown in the previous analysis, a window function can be applied to the integral and define intermediate domains between input and output. Applying the window function to the Fourier transform gives rise to the short-time Fourier transform, thus the intermediate domain is essentially a time-frequency joint distribution of the input data. Due to the uncertainty principle, a wider time-window leads to coarser timing resolution but finer frequency resolution, and vice versa. By varying the window size one can obtain a progressive transform between the time- and frequency-domains, factorizing the Fourier transform into incremental steps, which can be more efficiently mapped to a deep neural network than a brute-force fully-connected network model (Fig 10). We further notice that if the window function is widened by a factor of two at each hierarchical stage, the resulting data flow will resemble the classic radix-2 fast Fourier transform (FFT) algorithm, thus achieving a similar order of savings in terms computational complexity. It is conceivable that the hierarchical framework can be adapted to Fourier-related reconstructions such as MRI reconstruction. Overall, the extension and application of the hierarchical framework to a wide range of image reconstruction problems is possible.

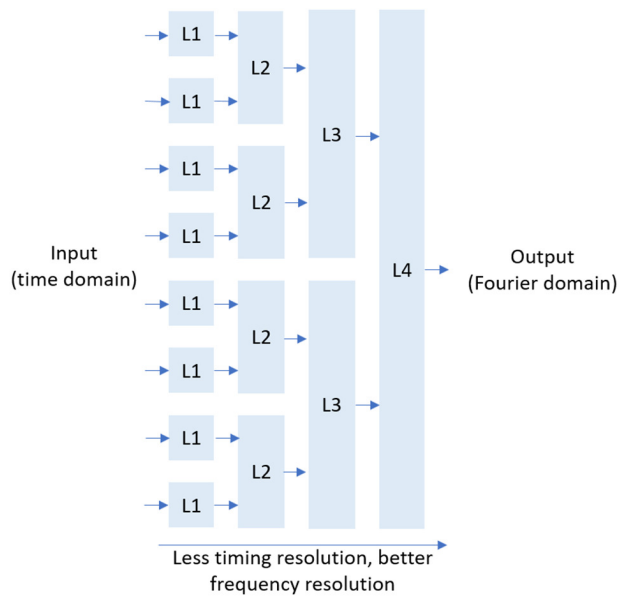

*Fig. 10. Illustration of the concept of hierarchical network for Fourier transform. Each block represents a sparse network. Overall the dataflow resembles the classic radix-2 fast Fourier transform.*

More broadly, the proposed methodology can be generalized to any transform or inverse problem, where some type of hierarchy can be defined explicitly (such as through Fredholm's equation) or implicitly (through end-to-end learning and using dynamic routing algorithms to prune out connections). Similarly, the approach can be used to solve large sets of equations and for numerical decomposition of large matrices.

## 5. Conclusions

We have presented a hierarchical approach to deep learning which enabled purely data-driven supervised-learning of CT reconstruction from full-size 2D data without relying on conventional analytical or iterative reconstruction algorithm structures. Sparse connection layers were introduced to implement the hierarchical network and reduce the dimensionality of the tomographic inversion problem. The network was partially trained with random noise patterns that encode the transform of interest. The image quality of these first learnt reconstruction results matches that of FBP reconstruction. In terms of computational cost, the hierarchical approach required only $O(n \log n)$ parameters compared to $O(n^2)$ parameters as needed by a generic network, making the proposed approach scalable to large data dimensions. In theory such a hierarchical approach should require a smaller order of arithmetic operations than analytical FBP reconstruction. The method opens the door to an entirely new type of reconstruction, which – with further improvements – could potentially lead to a new breakthrough in the tradeoff between image quality and computational complexity. The proposed hierarchical decomposition framework can be extended to Fourier transforms and other tomographic reconstruction problems. More broadly, it is conceivable to generalize the same methodology for solving any large-scale inverse problems and matrix decompositions.